\documentclass[prb,twocolumn,floatfix,nofootinbib]{revtex4}

\usepackage{graphics,epsf,times,amsfonts,bm,bbm}
\usepackage{latexsym}
\usepackage{amsthm}
\usepackage{amssymb}
\usepackage{amsmath}
\usepackage{graphicx}
\usepackage{hyperref}
\usepackage{hypernat}

\DeclareMathOperator{\DD}{DD}
\DeclareMathOperator{\CDD}{CDD}
\DeclareMathOperator{\PDD}{PDD}

\begin{document}

\title{High fidelity quantum memory via dynamical decoupling: theory and experiment}
\author{Xinhua Peng$^{1}$, Dieter Suter$^{1}$, Daniel A. Lidar$^{2}$}
\affiliation{$^{1}$Department of Physics, Technische Universit\"{a}t Dortmund, Germany \\
$^{2}$Departments of Electrical Engineering, Chemistry, and Physics, Center
for Quantum Information Science \& Technology, University of Southern
California, Los Angeles, CA 90089, USA}

\begin{abstract}
Quantum information processing requires
overcoming decoherence -- the loss of \textquotedblleft
quantumness\textquotedblright\ due to the inevitable interaction between the
quantum system and its environment. One approach towards a solution is
quantum dynamical decoupling -- a method employing strong and frequent
pulses applied to the qubits. Here we report on the first experimental test
of the concatenated dynamical decoupling (CDD) scheme, which invokes
recursively constructed pulse sequences. Using nuclear magnetic resonance,
we demonstrate a near order of magnitude improvement in the decay time of
stored quantum states. In conjunction with recent results on high fidelity
quantum gates using CDD, our results suggest that quantum dynamical
decoupling should be used as a first layer of defense against decoherence in
quantum information processing implementations, and can be a stand-alone solution in the
right parameter regime.
\end{abstract}

\maketitle

\section{Introduction}

One of the main difficulties in the implementation of quantum information processing
is the susceptibility of quantum systems to interactions with their environment.
Any such uncontrolled interaction degrades the quantum information stored in the system;
in the present context, this process is known as decoherence.\cite{Zurek} 
When a quantum system performs an information
processing task, decoherence causes
computational errors, which leads to the eventual loss of any quantum
advantage in information processing.\cite{Aharonov:96a} 
Dynamical decoupling (DD) is a form of quantum error \emph{suppression} that
modifies the system-environment interaction so that its overall effects are
very nearly self-canceling, thereby decoupling the system evolution from
that of the decoherence-inducing environment.\cite{Viola:99,Zanardi:98b} DD has
been primarily studied as a specialized quantum control technique for state 
preservation, or \textquotedblleft quantum memory\textquotedblright .
In this setting it has been demonstrated theoretically that DD\ is capable
of generating long-lived and robust quantum states. Three recently
introduced techniques have figured most prominently in this effort:
\textquotedblleft randomized DD\textquotedblright\ (RDD),\cite{Viola:05}
\textquotedblleft concatenated DD\textquotedblright\ (CDD),\cite%
{KhodjastehLidar:04} and \textquotedblleft Uhrig DD\textquotedblright\ (UDD).%
\cite{Uhrig:07} RDD, which involves randomly selected pulse types
applied at regular intervals, works best for strongly time-dependent environments and
in the long time limit.\cite{Santos:06,Zhang:08} UDD, which involves
an optimization of pulse intervals for a fixed pulse type, is provably optimal for
the restricted case of a single qubit with diagonal coupling to the
environment.\cite{Yang:08,Pasini:09} UDD\ has been the subject of extensive
recent experimental tests. Following some notable successes,\cite%
{Biercuk:09,biercuk-2009,Jenista2009Optimized} the general picture emerging
is that UDD performance depends sensitively on the existence of a sharp
high-frequency cutoff; in the absence of such a cutoff simpler DD\ schemes
tend to result in better performance.\cite%
{Ryan2010Extending,Szwer2010Keeping,Ajoy2010Optimal} UDD was recently
extended to \textquotedblleft quadratic DD\textquotedblright\ (QDD) which
applies to general environments, but is still limited to only a single
qubit.\cite{WFL:09} Very recently a generalization of QDD to multiple qubits
was proposed, 
which involves a higher order nesting of UDD sequences.\cite{Wang:10}
CDD is a method for constructing pulse sequences with a recursive
structure. While
CDD requires longer pulse sequences than UDD and QDD, there is strong
theoretical evidence that it can be successfully combined with quantum
computation and incorporated into quantum logic circuits.\cite{NLP:09,West:10} For this reason we focus here on CDD, and present the first
experimental evidence,
using nuclear magnetic resonance (NMR)\ techniques, for the theoretically
predicted advantage\cite%
{KhodjastehLidar:04,KhodjastehLidar:07,Witzel:07a,PhysRevB.75.201302,Zhang:08,NLP:09,West:10}
of using CDD\ for quantum memory preservation over its common counterpart
\textquotedblleft periodic dynamical decoupling\textquotedblright\
(PDD).\footnote{Since this work first appeared as a preprint
  (arXiv:0911.2398)\ a number of other experimental studies comparing
  CDD\ and PDD\ have been published
  \cite{Alvarez:10,2010arXiv1011.1903T,2010arXiv1011.6417W,2010arXiv1007.4255B}.}
We also provide a theoretical analysis in support of these
experimental results. Though our experimental results illustrate the
effectiveness of CDD specifically for NMR, the CDD framework is
generally applicable to any decohering quantum system. Our results
lend support to the expectation that quantum dynamical decoupling will
prove an indispensable tool in scalable quantum information
processing.

\section{Dynamical decoupling and CDD}

We consider a system described by a Hamiltonian 
\begin{equation}
H = H_{S} + H_{B} + H_{SB},
\end{equation}%
where $H_{S}$ acts on the system degrees of freedom, $H_{B}$ on the environment
and $H_{SB}$ couples the system to the environment. 
To suppress errors, DD allows the system to evolve for some time before
applying a control pulse to redirect or refocus the evolution toward the
error-free ideal, continually repeating this process until some total
evolution has completed: 
\begin{equation}
\DD[U(\tau)]=P_{N}U(\tau _{0})\cdots P_{2}U(\tau _{0})P_{1}U(\tau _{0})=%
\widetilde{U}(N\tau _{0}),
\end{equation}%
where $U(\tau _{0})=U_{0}(\tau _{0})B(\tau _{0})$ represents the unitary
evolution of the combined system and environment, or bath, for a duration of
length $\tau _{0}$, decomposed so that $U_{0}(\tau _{0})$ determines the
ideal, desired error-free evolution generated by a piece-wise constant
system Hamiltonian $H_{S}$, and $B(\tau _{0})$ is a unitary error operator
acting jointly on the system and bath. We use the symbol
$\widetilde{U}$ to denote evolution in the presence of DD pulses. For now, we implicitly assume that
the pulses $P_{j}$ are sufficiently fast as to not contribute to the total
time of the evolution. This assumption is not essential, but is useful in
simplifying our present discussion. The simplest example is quantum memory,
where $U_{0}(\tau _{0})=I_{S}$ is the identity operation or
\textquotedblleft free-evolution\textquotedblright\ of the system,\ and $%
B(\tau _{0})$ represents the deviation from the ideal dynamics caused by the
presence of a bath. In this case, our goal is to choose pulses so that $%
\DD[U(\tau)]=I_{S}\otimes \widetilde{B}$, the identity acting on the system
and an arbitrary pure bath operator $\widetilde{B}$. DD schemes differ in
precisely how these pulses $P_{j}$ are chosen, with the only common
constraint that the following basic \textquotedblleft decoupling
condition\textquotedblright\ is met:\cite{Viola:99,Zanardi:98b} 
\begin{equation}
\sum_{\alpha }P_{\alpha }^{\dag }H_{SB}P_{\alpha }=0.  \label{basicdd}
\end{equation}%
To be concrete, we will suppose that the pulses $P_{\alpha }\in \{I,X,Y,Z\}$
are Pauli operators, though again this assumption only serves to keep the
explanation simple and is not strictly necessary. CDD generates pulse
sequences by \emph{recursively} building on a base sequence $Z[\cdot
]X[\cdot ]Z[\cdot ]X[\cdot ]$ (or any other ordered pair of Pauli
operators from the set $\{X,Y,Z\}$), as follows. The sequence is initialized as 
\begin{equation}
\CDD_{0}[U(\tau _{0})]=U(\tau _{0})=U_{0}(\tau _{0})B(\tau _{0})\equiv 
\widetilde{U}_{0}(\tau _{0}),
\label{CDD0}
\end{equation}%
and higher levels are generated via the rule
\begin{eqnarray}
&&\widetilde{U}_{n+1}(\tau _{n+1})\equiv \CDD_{n+1}[U(\tau _{0})] =\notag \\
&&Z[\widetilde{U}_{n}(\tau _{n})]X[\widetilde{U}_{n}(\tau _{n})]Z[\widetilde{U}_{n}(\tau _{n})]X[\widetilde{U}_{n}(\tau
_{n})],
\label{eq:CDD-def}
\end{eqnarray}
where $\tau _{n}=4^{n}\tau _{0}$. Thus, at each level the total duration of
the CDD\ pulse sequence grows by a factor of $4$, while the pulse interval
remains fixed at its initial value $\tau _{0}$. Note that we are allowing
for the possibility of some non-trivial information processing operation $%
U_{0}(\tau _{0})$, as implemented in Ref.~\onlinecite{West:10}. The choice of the
base sequence is motivated by the observation that it satisfies the
\textquotedblleft decoupling condition\textquotedblright\ (\ref{basicdd}) in
the quantum memory setting $U_{0}(\tau _{0})=I_{S}$, under the dominant
\textquotedblleft 1-local\textquotedblright\ system-bath coupling term $%
H_{SB}^{(1)}=\sum_{\alpha =x,y,z}\sum_{j}\sigma _{j}^{\alpha }\otimes
B_{j}^{\alpha }$, where $\sigma _{j}^{x}\equiv X$, $\sigma _{j}^{y}\equiv Y$%
, and $\sigma _{j}^{z}\equiv Z$ denote the Pauli matrices acting on system
qubit $j$, and $\{B_{j}^{\alpha }\}$ are arbitrary bath operators. (The next
order \textquotedblleft 2-local\textquotedblright\ coupling would have terms
such as $\sigma _{j}^{\alpha }\sigma _{k}^{\beta }\otimes B_{jk}^{\alpha
\beta }$, etc.) For this reason the base sequence is sometimes called the
\textquotedblleft universal decoupler\textquotedblright . Similarly, the
most common pulse sequence used thus far in DD\ experiments is
\textquotedblleft periodic DD\textquotedblright\ (PDD), which generates
pulse sequences by \emph{periodically} repeating the universal decoupler
base sequence $Z[\cdot ]X[\cdot ]Z[\cdot ]X[\cdot ]$: 
\begin{equation}
\PDD_{k}[U(\tau _{0})]=(\PDD_{1}[U(\tau _{0})])^{k}=\widetilde{U}_{k}(4k\tau
_{0}),  \label{eq:PDD-def}
\end{equation}%
where $\PDD_{1}[U(\tau _{0})]=\CDD_{1}[U(\tau _{0})]$. Note that in
Eq.~(\ref{eq:PDD-def}) $\widetilde{U}_{k}$ refers to the $k$-fold
\emph{repetition} of single cycles of the sequence, while in
Eq.~(\ref{eq:CDD-def}) $\widetilde{U}_{n}$ describes the $n$-fold \emph{concatenation} of identical cycles.

\begin{figure}[tph]
\centering 
\includegraphics[width=3.2in]{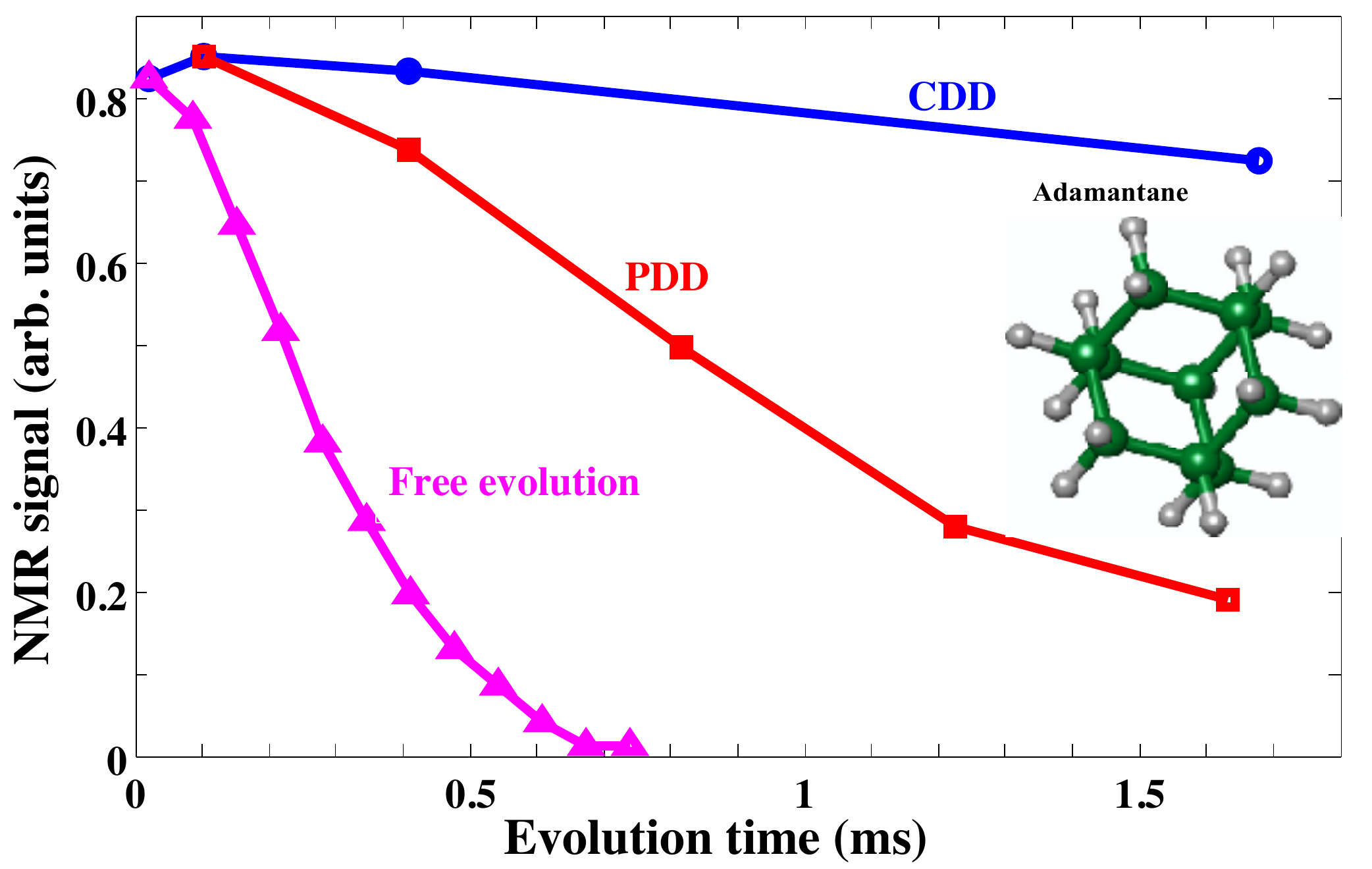}
\caption{Experimental comparison of CDD vs PDD vs free evolution on a $300$MHz NMR
spectrometer.  Shown is the signal decay of the ${^{13}}$C-spin
coherence of Adamantane (C$_{10}$H$_{16}$, depicted in the inset)
vs. total evolution time.  The spin echo is measured by first creating
the equal superposition state $(|0\rangle +|1\rangle
)/\protect\sqrt{2}$, then subjecting it to either PDD (red squares) or
CDD (blue circles) pulse sequences, and finally measuring the free
induction decay.  The first point shows the amplitude of the free
induction decay (FID) signal, measured after a pulse interval of
$\protect\tau_0 =15\protect\mu s$ [i.e., $\CDD_0$, Eq.~(\ref{CDD0})].  The second point corresponds to
$\CDD_{1}=\PDD$.  The third and the fourth points correspond to
$\CDD_2$ and $\CDD_3$, respectively.  The width of the $\pi$-pulses
was $\protect\delta =10.52\protect\mu s$. For reference we also show
the free evolution, without any pulses (triangles).}
\label{nmrsignal}
\end{figure}

The use of periodic pulse sequences has a long history, starting with some
of the earliest spin-echo experiments \cite{Carr:54}. Indeed, it is natural
to try to design a \textquotedblleft good\textquotedblright\ pulse sequence
and then repeat it over and over again so as to cover a desired total
experiment time $T$. CDD breaks with this intuition and tradition, by
demanding that a different pulse sequence be used for every given $T$:\
given a feasible pulse interval $\tau _{0}$, the appropriate CDD\ order $n$
is found from the condition $T=4^{n}\tau _{0}$. Unlike in PDD, the pulse
order in CDD level $n+1$ is completely different from that in level $n$.

Next we present
the results of the first-ever CDD experiment.\footnote{These
  experiments were first carried out in May 2005 in collaboration with
  Marko Lovric, and subsequently refined and
  presented in various talks given by the authors.} The results demonstrate that,
as predicted,\cite%
{KhodjastehLidar:04,KhodjastehLidar:07,Witzel:07a,PhysRevB.75.201302,Zhang:08,NLP:09,West:10}
CDD\ outperforms PDD in preserving quantum memory, and CDD\ becomes
increasingly effective at higher concatenation levels.

\section{Experimental results on quantum memory using CDD}

\begin{figure}[tph]
\centering 
\includegraphics[width=3.2in]{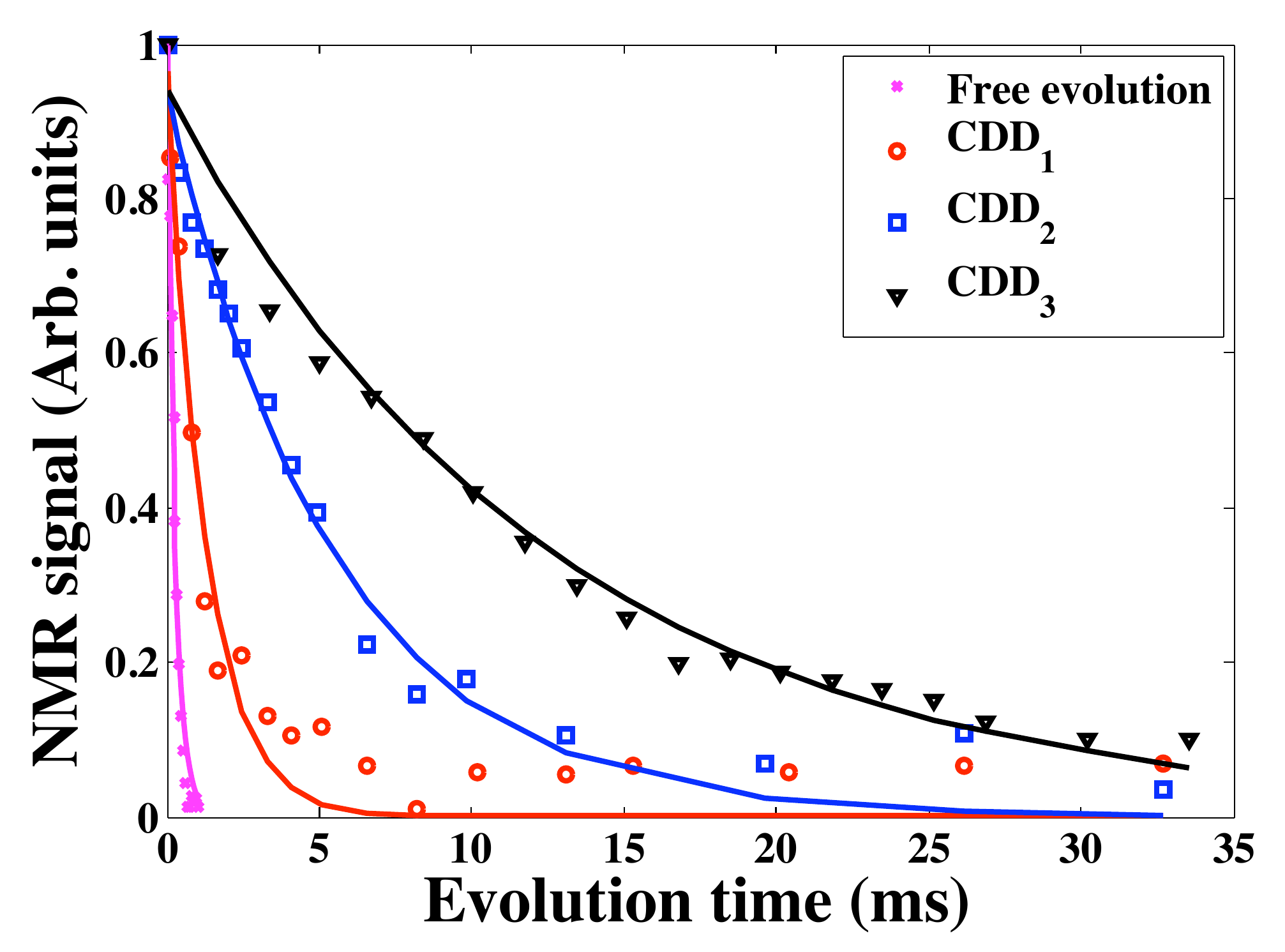} 
\caption{Signal decays of $^{13}$C-Spin for the free evolution,
CDD$_1$, CDD$_2$ and CDD$_3$ with $\tau_0 = 15 \mu s$. The solid lines
denote the fitting to an exponential decay.}
\label{signaldecay}
\end{figure}

\begin{figure}[tph]
\centering 
\includegraphics[width=3.2in]{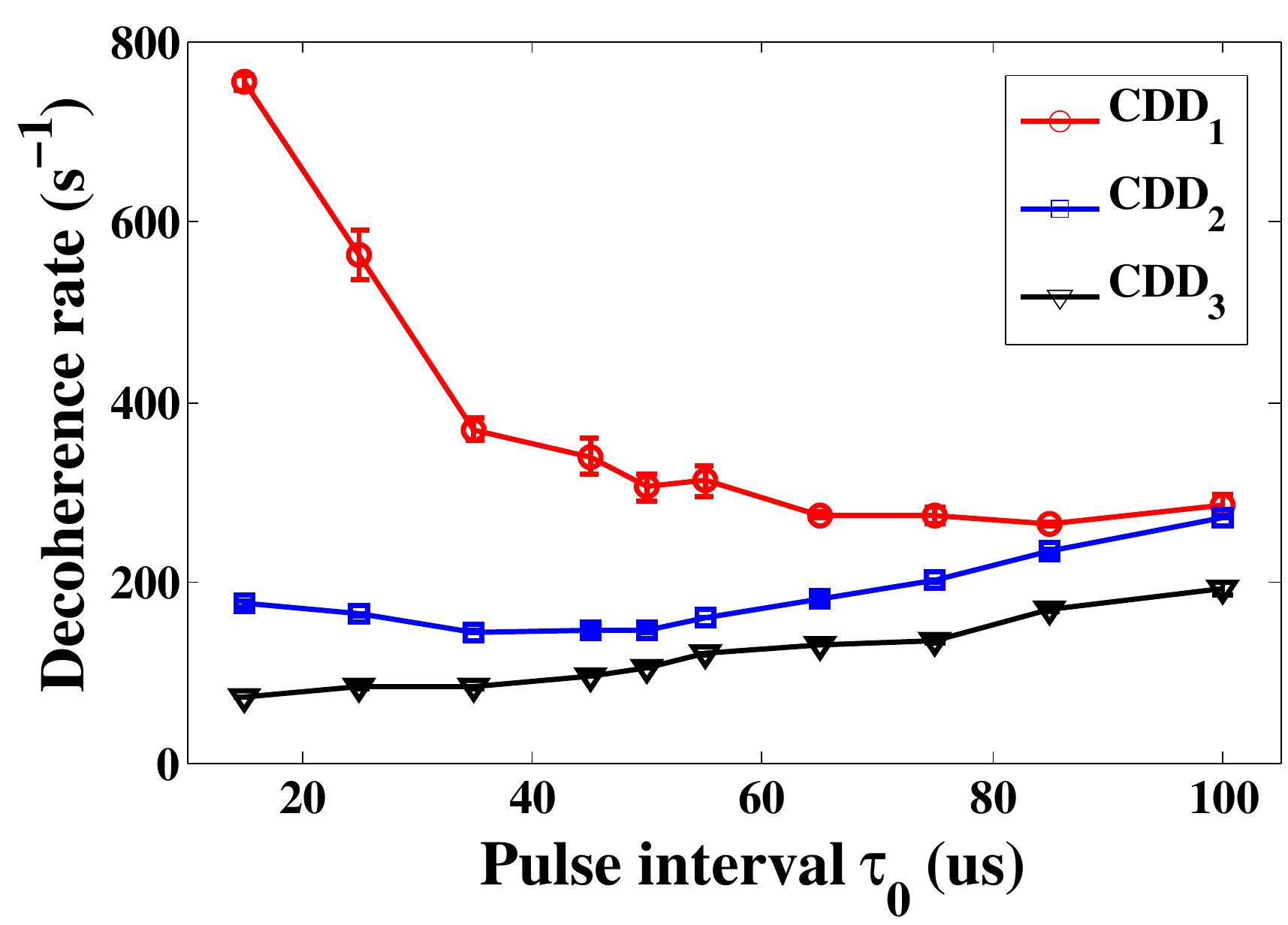} 
\caption{Decoherence rate vs. pulse interval for the system shown in
Figure~\protect\ref{nmrsignal}, for $\CDD_{1}=\PDD_{1}$ (red,
circles), $\CDD_{2}$ (blue, squares), and $\CDD_{3}$ (black,
triangles). 
For PDD with non-ideal pulses, errors due to
finite pulse-width accumulate and overwhelm the improvement due to the
smaller pulse interval. CDD, on the other hand, compensates for finite
pulse width errors, and hence improvement is seen with smaller pulse
intervals, as expected.}
\label{t2}
\end{figure}

Presented in Figures~\ref{nmrsignal}-\ref{t2} are the results of quantum
memory NMR spectroscopy experiments. The experiments were performed at TU
Dortmund with a homebuilt 300 MHz solid-state NMR spectrometer using a 7.05
T Oxford magnet and a homebuilt double resonance (HC) probe. Before the
experiments, we first used suitable tune-up pulse sequences \cite%
{Mehring:72,Haubenreisser:79,Burum:81} to minimize pulse-length errors and
phase transients for the $^{13}$C channel by adjusting the probe tuning and
RF amplitudes. For the tune-up, we used a liquid sample of $^{13}$C-labeled
Methanol (99\% enriched).

To measure the effect of the decoupling sequences, we used a sample of
powdered Adamantane. In this plastic crystal, the nearly spherical molecules
tumble rapidly and isotropically in the solid phase. The motion averages all
intramolecular dipolar couplings to zero, but does not eliminate
intermolecular couplings. As a result of this averaging process, there is
only one coupling between every pair of molecules, thereby reducing the
Adamantane molecule to a point dipole source containing 16 proton spins or
16 proton spins and one $^{13}$C-nuclear spin in any C position (11\% of the
molecules). All remaining interactions (except for the Zeeman term) are
insignificant. The sensitivity of $^{13}$C NMR signals was enhanced by the
standard cross-polarization experiment, which provides polarization transfer
from the abundant $^1$H spins. The enhancement factor of the $^{13}$C signal
was about 3.2. After the transfer, the carbon spin magnetization was stored
for 2 ms as longitudinal magnetization. Another $\pi$/2 pulse created
transverse $^{13}$C magnetization as the initial state for the dynamic
decoupling experiments.

In the experiments, the width $\delta $ of the $\pi $ pulses for $^{13}$C
was 10.52 $\mu $s. Pairs of $\pi $ pulses with the same phase and no
intervening delay were omitted. For pairs of $\pi $ pulses with different
phases, we inserted a minimum delay $f_{a}=376$ ns to change the phase of
the rf-pulse. As an example, $XZf$ (where $f$ denotes free evolution,
without pulses) was implemented as $Xf_{a}Zf$. The minimum cycle time for
the PDD sequence was $\tau _{c}^{p_{1}}=4(\tau _{0}+\delta )$, for the
even-order CDD sequences $\tau _{c}^{p_{2k}}=4\tau _{c}^{p_{2k-1}}+2f_{a}$,
and for the odd-order CDD sequences $\tau _{c}^{p_{2k+1}}=4(\tau
_{c}^{p_{2k}}+\delta) $, where $\tau _{0}$ is the interval between
consecutive $\pi $-pulses. The remaining signal was detected by measuring
the free induction decay of the $^{13}$C spins while decoupling the protons
after the last echo pulse and a dead-time of $\tau _{d}=8.5\mu $s. To obtain
an accurate signal amplitude, we Fourier-transformed the free induction
decays and integrated the signals over the relevant frequency range.

Figure~\ref{nmrsignal} shows the spin echo signal for the $^{13}$C 
nuclear spin qubits.
 In each experiment, we applied cyclic sequences of $\pi$-pulses  corresponding to
 the various CDD orders and measured the remaining signal after the end of the sequence.
We plot the absolute value of the transverse magnetization, corresponding to
\begin{equation}
 \vert \langle \sigma _x - i \sigma _y \rangle \vert  =
 \sqrt{{\rm Tr}[\sigma _x \rho(\tau _{c}^{p_{k}})]^2 +  {\rm Tr}[\sigma _y \rho(\tau _{c}^{p_{k}})]^2 },
\end{equation}
 where the density matrix $\rho(\tau _{c}^{p_{k}})$
 represents the state of the system at the end of DD pulse sequence
 (either PDD or CDD, where $k$ now represents the number of cycle
 repetitions (for PDD) or number of concatenations (for CDD)).
Figure ~\ref{signaldecay} shows the decay of the signal as the number of cycles and thus the total evolution time increases.
In these experiments, the delay between pulses was set to $\tau_0 = 15 \mu s$.
By fitting the experimental data to an exponential decay function $S = S_0 e^{-t/T_2}$, 
we extracted the decoherence rates $1/T_2$. 
Figure~\ref{t2} shows
the decoherence rates extracted from a series of experiments conducted with
different pulse intervals $\tau _{0}$.

 These data show that: (i) higher
levels of CDD suppress errors more efficiently than lower levels, (ii) at
high concatenation levels the benefits of CDD increase as $\tau _{0}$
decreases, (iii) CDD always and substantially outperforms PDD, an
improvement that becomes more pronounced as $\tau _{0}$ decreases. Indeed,
comparing the results for $\CDD_1=\PDD$ and $\CDD_3$ at $\tau
_{0}=15\mu s$ in Fig.~\ref{t2}, we observe an improvement of almost an order of
magnitude in the decoherence rate. 

We note the decoherence here is not
engineered, but instead caused by the actual dipole-dipole interactions
between the $^{13}$C qubit and the $^{1}$H spins inherent in the Adamantane
molecule. 
We further note that the specific nature of the interaction, be it
dipole-dipole, Heisenberg, or something else altogether, is inconsequential
to the success of CDD. What matters is the strength of this interaction
relative to the DD\ pulse interval $\tau _{0}$, as will be made more
explicit in the theoretical analysis, to which we now turn.

\section{Theoretical analysis}

We now present a theoretical interpretation of some of the results shown in
Figures \ref{nmrsignal}-\ref{t2}. To allow for a generic analysis of the
DD response to different bath environments and system-bath couplings, we
characterize the leading-order DD behavior for simplicity in terms of the
following relevant parameters, which capture the strength or overall rate of
the internal bath and system-bath dynamics, respectively: $\beta \equiv
\left\Vert H_{B}\right\Vert $ and $J\equiv \left\Vert H_{SB}\right\Vert $.\cite{UCDD:com2}
The total Hamiltonian is $H=H_{S}+H_{B}+H_{SB}$. If $J\gg \beta $, then the system-bath coupling is a
dominant source of error, and remains relatively stable since the internal
bath dynamics are comparatively slow. In this case, DD should produce
significant fidelity gains as it removes the dominant error source. On the
other hand, if $J<\beta $, then the system coupling to the environment
induces relatively slow dynamics, while the environment itself has fast
internal dynamics. In this case, minimizing the system-bath coupling will
have less of an effect on the overall dynamics, so it may be considered a
worst case scenario when assessing DD performance.

Let us now compare the pure system state $\rho _{S}^{0}$ obtained in the
error-free setting, where $H_{SB}=0$ (the desired state), to the system
state $\rho _{S}$ resulting from the presence of a non-zero $H_{SB}$, but
subject to CDD level $n$ (the actual state). The appropriate distance
measure, which we would like to minimize, is the trace-norm distance $D$,
defined in Appendix~\ref{app:A}. For a total evolution time $T=N\tau _{0}$
using $N=4^{n}$ zero-width pulses, the distance $D$ between the desired and
actual state is bounded, in the \textquotedblleft
pessimistic\textquotedblright\ regime $J<\beta $, as:\cite%
{KhodjastehLidar:07,LZK:08,NLP:09} 
\begin{equation}
D[\rho _{S}(T),\rho _{S}^{0}(T)]\lesssim 2J\tau _{0}\epsilon ^{n},
\label{eq:D}
\end{equation}%
where $\epsilon \equiv 4\beta \tau _{0}2^{n}$. Here $\epsilon $ plays the
role of a \textquotedblleft threshold parameter\textquotedblright , i.e.,
rapid convergence of the distance to zero in the large $n$ limit is
guaranteed if $\epsilon <1$. This situation is reminiscent of the error
threshold in quantum fault tolerance theory,\cite{Aliferis:05} but requires us to decrease $\tau _{0}$ while the
concatenation level rises, in order to mitigate the $2^{n}$
factor. In general this is
undesirable since the pulse interval cannot be decreased
indefinitely. Indeed, in each of our experiments the pulse interval
was fixed and the total time was variable. The interval length was
varied from experiment to experiment, and the total time was adjusted accordingly, depending on the number of pulses used.
Therefore we next analyze the above bound on $D$ under the assumption
of fixed pulse interval $\tau _{0}$. We shall show that, perhaps
counterintuitively, there is a regime where increasing the
concatenation level, and hence also the total evolution time, actually
leads to improved fidelity. Such an improvement can, however, not
continue indefinitely, and we shall find that there is an optimal
concatenation level beyond which fidelity starts to drop.

If $\delta ^{\ast }$ is the desired maximum probability of distinguishing $%
\rho _{S}(T)$ from $\rho _{S}^{0}(T)$, we can set the bound (\ref{eq:D}) on $D$
equal to $\delta ^{\ast }$. This yields a quadratic equation in the number
of concatenation levels $n$, whose physically meaningful branch is:%
\begin{eqnarray}
n=&-&\hspace{-.2cm}\sqrt{[1+\log _{4}\!(\beta \tau _{0})]^{2}-2\log _{4}\!({J\tau
    _{0}}/{\delta ^{\ast }})-1}\notag \\
&-&\hspace{-.2cm}[1+\log _{4}(\beta \tau _{0})].
\end{eqnarray}
(Rounding to the nearest integer is implied.) This quantifies the \textquotedblleft simulation overhead\textquotedblright\
in CDD (the analog of the 
poly-logarithmic simulation overhead in fault tolerance theory\cite{Aliferis:05}), and means that using CDD, in
the limit of zero-width pulses, we can obtain a given simulation accuracy
target $\delta ^{\ast }$ with a number of concatenation levels that scales
only \emph{logarithmically} in $\beta \tau _{0}$, $J\tau _{0}$, and $\delta
^{\ast }$. We can also minimize the bound (\ref{eq:D}) on $D$ for $n$ at
fixed $J\tau _{0}$ and $\beta \tau _{0}$. The result is%
\begin{equation}
n_{\mathrm{opt}}= \lfloor \log _{4}\frac{1}{\beta \tau _{0}}-1
\rfloor,
\label{nopt}
\end{equation}%
which is the optimal number of concatenation levels, in agreement with more
involved derivations.\cite{KhodjastehLidar:07,NLP:09} At this optimal
value, which increases with decreasing $\beta \tau _{0}$, the bound on $D$
scales as $e^{-\frac{1}{2}(\log _{4}\beta \tau _{0})^{2}}$ and also becomes
very small in the small $\beta \tau _{0}$ limit. Since the internal bath
dynamics are the dominant error source here, it is not surprising that the
optimal concatenation level only depends on $\beta \tau _{0}$. A similar
expression for optimal concatenation level occurs in the \textquotedblleft
optimistic\textquotedblright\ $J>\beta $ regime, depending on both $J$ and $%
\beta $.\cite{KhodjastehLidar:07,NLP:09} The key point is that even in the simplest case: ideal,
single-qubit quantum memory, there exists an optimal concatenation level
depending on the $J\tau _{0}$ and $\beta \tau _{0}$ parameters, after which
point increasing concatenation no longer improves system fidelity.
Intuitively, this optimal concatenation level represents the turning point
between the competing processes of improved CDD error suppression and the
exponentially increasing time for error-prone evolution that occurs with
each level of concatenation. The precise point at which DD error suppression
is overcome by the accumulated error depends on the relative strengths of
the error-inducing Hamiltonians, determined by $J\tau _{0}$ and $\beta \tau
_{0}$. For comparison, let us also consider the case of PDD with $N$ pulses.
In this case the bound $D[\rho _{S}(T),\rho _{S}^{0}(T)]\lesssim 2NJ\tau
_{0}\beta \tau _{0}$ applies under the same assumptions as those required
for the CDD bound (\ref{eq:D}) on $D$. The evidently much more rapid
convergence of CDD, as long as pulse intervals shorter than $1/(2^{n+2}\beta
)$ are attainable (the $\epsilon <1$ condition), is due to its recursive
structure, which allows CDD to cancel successive orders of the
system-bath propagator in time-dependent perturbation theory.\cite{KhodjastehLidar:07,NLP:09}

Returning now to our experimental CDD\ results (Fig.~\ref{t2}), which are in the
\textquotedblleft pessimistic\textquotedblright\ regime $J<\beta $,\cite{UCDD:com3} we see that for $\CDD_3$ they agree with
the theoretical prediction that under concatenation, lowering $\tau _{0}$ improves the
fidelity.\cite{UCDD:com1}
Moreover, they are consistent with the $\epsilon <1
$ regime, where CDD\ outperforms PDD. 
However, as can be seen from Figure~\ref{nmrsignal}, while CDD
consistently outperforms PDD, and both are better than free
(unprotected) evolution, the performance of both PDD and CDD
deteriorates with increasing total experiment time $T$. Thus, our
experiments have $n_{\rm opt}=1$ in the
sense of the optimal concatenation level of
Eq.~(\ref{nopt}). In other words, our experiments did not reach the
``sweet spot'' where
a higher level of concatenation (and hence total evolution time)
actually leads to increased fidelity, which would lead to $n_{\rm opt}>1$.
The closeness of $\epsilon $ to
$1$ and the finiteness of the pulse width play a role in this regard.

\section{Summary and Conclusions}

Early DD schemes were designed to remove unwanted system-bath
interactions to a given, low order in time-dependent perturbation
theory \cite{Viola:99,Zanardi:98b}. CDD is the first explicit scheme proposed to
remove such interactions to an arbitrary order, via a recursive
construction in which each successive level removes another order in
time-dependent perturbation theory \cite{KhodjastehLidar:04}. This
work reports the first experimental demonstration of CDD, using NMR. As predicted,
CDD outperforms PDD in preserving the fidelity of quantum states. The
experimental system studied here---the $^{13}$C spin qubit of
Adamantane---undergoes single-axis decoherence, so that one might
suspect that the full CDD
sequence we used is ``overkill'' for this system, because this sequence was
designed to suppress full three-axis decoherence. However, pulse
errors and magnetic field inhomogeneity result in second order
off-axis contributions to decoherence, which is why DD sequences using
just a single pulse type (e.g., UDD), cannot be expected to perform
well on our system. Nevertheless, it is important to identify an
experimentally feasible NMR system where dephasing and longitudinal
relaxation are of comparable magnitude, and to study the performance
of CDD relative to other methods capable of suppressing full
decoherence, such as QDD. 

\acknowledgments This work is supported by the DFG through Su
192/24-1. D.A.L. was sponsored by the NSF under Grants No. CHM-924318 and
CHM-1037992.

\appendix

\section{Distance and fidelity}

\label{app:A} A distance measure between quantum states $\rho $ and $\sigma $
(density matrices) is provided by the ``trace-norm distance'': $D(\rho
,\sigma )\equiv \frac{1}{2}\Vert \rho -\sigma \Vert _{1}$, where $\Vert
A\Vert _{1}\equiv \mathrm{Tr}\sqrt{A^{\dagger }A}=\sum_{i}s_{i}(A)$, and
where $s_{i}(A)$ are the eigenvalues of $\sqrt{A^{\dagger }A}$. The trace
norm distance is the maximum probability of distinguishing $\rho $ from $%
\sigma $, vanishes if and only if $\rho =\sigma $, and is related to the
fidelity\ $F(\rho ,\sigma )\equiv \Vert \sqrt{\rho }\sqrt{\sigma }\Vert
_{1}\,$via $1-D(\rho ,\sigma )\leq F(\rho ,\sigma )\leq \sqrt{1-D(\rho
,\sigma )^{2}}$, so that knowing one bounds the other. In the case of
comparing an output state $\sigma $ to a desired pure state $\rho
=\left\vert \psi\right\rangle\left\langle \psi\right\vert$, this reduces to $%
F=\sqrt{\left\langle \psi\right\vert\sigma \left\vert \psi\right\rangle}$,
and perfect fidelity ($F=1 $) results if and only if $\rho =\sigma $.


\begin{thebibliography}{38}
\bibitem[{UCD({\natexlab{a}})}]{UCDD:com2}
\bibinfo{note}{We note that the bounds below, which involve the operator norms
  $J$ and $\beta $, are undoubtedly overly pessimistic, and should be replaced
  by correlation functions in a more careful analysis. See, e.g.,
  Ref.~\onlinecite{NLP:09}.}
\bibitem[{UCD({\natexlab{b}})}]{UCDD:com3}
\bibinfo{note}{We estimate $J\sim 600$Hz and $\beta \sim 2.5$kHz on the basis
  of a simulation: a spin system consisting of one $^{13}$C and six $^{1}$H
  with these parameters produced the best fit to the experimental free
  induction decay.}
\bibitem[{UCD({\natexlab{c}})}]{UCDD:com1}
\bibinfo{note}{We are taking some liberty here with the term ``fidelity'': our
  theoretical analysis indeed provides a fidelity bound, while the experimental
  results report different, but related quantities, namely NMR signal and
  decoherence rate. Moreover, our experiment is restricted to testing the
  performance of CDD for a single input state, whereas the theoretical distance
  and fidelity measures we have discussed apply for general input states.}
\end{thebibliography}

\end{document}